BMC Medical Informatics and
Decision Making

# Investigating usability of mobile health applications in Bangladesh

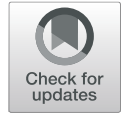

Muhammad Nazrul Islam[1*] , Md. Mahboob Karim[1], Toki Tahmid Inan[1] and A. K. M. Najmul Islam[2]

## Abstract

**Background:** Lack of usability can be a major barrier for the rapid adoption of mobile services. Therefore, the purpose of this paper is to investigate the usability of Mobile Health applications in Bangladesh.

**Method:** We followed a 3-stage approach in our research. First, we conducted a keyword-based application search in the popular app stores. We followed the affinity diagram approach and clustered the found applications into nine groups. Second, we randomly selected four apps from each group (36 apps in total) and conducted a heuristic evaluation. Finally, we selected the highest downloaded app from each group and conducted user studies with 30 participants.

**Results:** We found 61% usability problems are catastrophe or major in nature from heuristic inspection. The most (21%) violated heuristic is aesthetic and minimalist design. The user studies revealed low System Usability Scale (SUS) scores for those apps that had a high number of usability problems based on the heuristic evaluation. Thus, the results of heuristic evaluation and user studies complement each other.

**Conclusion:** Overall, the findings suggest that the usability of the mobile health apps in Bangladesh is not satisfactory in general and could be a potential barrier for wider adoption of mobile health services.

## Background

Digitalization can play an important role in delivering health services to individuals and communities. Hospitals have been digitalizing their services and renovating the whole process of health care. Digitalization not only helps to improve patient safety and satisfaction but also keeps health statistics of the population up-to-date. Since an approximate of 6 billion people (around 75% of the world population) have access to mobile phones, mobile health applications have become an important channel for providing healthcare services by healthcare service providers [1]. Therefore, mHealth (mobile health) as a field of research has become prominent in recent years. In general, the term mHealth refers to the practice of using mobile devices in health care services [2]. The Global Observatory for eHealth defines mHealth as "*medical and public health practice supported by mobile devices, such as*

*mobile phones, patient monitoring devices, personal digital assistants (PDAs), and other wireless devices*" [3]. The mHealth applications include the use of mobile devices to improve the process of medical data collection [4], service delivery [5], patient-doctor communication [6], and real-time monitoring and adherence support [7].

The mHealth applications have even been widely used in the remote villages of developing countries [8]. For example, doctors in Tanzania can use *Afya Mtandao* (Swahili for Health Network) network from anywhere in the country even during the surgical operation through mobile phones [9]. United States Agency for International Development (USAID) has built the foundation of Mobile Alliance for Maternal Action (MAMA), a support for the development of a health informatics network for maternal patients using mobile phones. Monitoring and Evaluation to Assess and Use Results (MEASURE) is another US (United States) funded project that focuses mainly on the development of mobile platforms for monitoring health in developing countries

---
* Correspondence: nazrul@cse.mist.ac.bd
[1]Department of Computer Science and Engineering, Military Institute of Science and Technology, Dhaka 1216, Bangladesh
Full list of author information is available at the end of the article

[10]. Most of these mHealth services in developing countries rely on text messaging [10] in health related issues such as HIV (human immunodeficiency virus) prevention [[11–14]]. These messages are also used to bridge the community and the health worker [15].

In Bangladesh, a number of initiatives on mHealth have already been taken by the government and the non-government organizations. These include developing mobile applications for health care service delivery, personal health tracking, remote consultation, and information delivery. However, the outcomes of these initiatives are not well understood. There has been very little research in the mHealth domain of Bangladesh. There are a few studies, which mainly focused to assess the opportunities, and challenges of using Information Technology in the health sector of Bangladesh [16–19]. However, very few studies have been conducted that evaluate the mHealth applications in Bangladesh in terms of usability. Usability is defined as the effectiveness, efficiency, and satisfaction with which users achieve specified goals in a specific context of use [20]. Usability is one of the key quality attributes for the successful development and adoption of mHealth apps like any other digital applications [21–24] . Therefore, the objective of this research is to understand the state of the art of the mHealth applications developed in Bangladesh, and to assess the overall usability of these applications. Consequently, this paper addresses the following research questions:

*How usable are the mHealth applications in Bangladesh? Does the usability vary depending on the type of applications?*

To attain this research objective, this paper follows a 3-stage research process: systematic application review, expert inspection, and user studies. The findings suggest that the usability of the mHealth apps in Bangladesh is not satisfactory in general. Based on the findings, the paper provides several actionable guidelines for the practitioners.

ICT (Information and Communication Technology) in healthcare services has brought revolutionary changes by enhancing service quality, ensuring safety and clinical effectiveness. Research on health informatics, in the form of mHealth, has gained much attention [25]. A synopsis on the impact of mobile applications in public health can be found in the study of Fiordelli et al. [25]. They reviewed mHealth related studies published between 2002 and 2012 and found that mHealth services are mostly focused on chronic conditions. They also found a lack of mHealth related studies in Asia. In fact, very few studies have been conducted on the influence and adequacy of mHealth services in developing countries like Bangladesh.

In order to understand the influence of mHealth on public health services, Ahmed et al. [19] studied the existing eHealth and mHealth enterprises and assessed their prospects in Bangladesh. They found that tele-consultation, prescription, and referral are the most common initiatives. They also suggested that the availability of trained health care professionals, capacity building, research support, and experience factors sharing are the key for promoting and implementing mHealth services in Bangladesh. In another study, Prodhan et al. [26] investigated the telemedicine initiatives and their interoperability. They found that most of the initiatives have different formats of data. Different data format along with lack of proper data storage facilities greatly affect the interoperability among the existing health care services. They further found that most of these projects work well at the beginning, but fail in the long run due to lack of publicity and user acceptance. Thus, they recommend for a standard data storage policy as well as a proper awareness program to increase the usage of telemedicine services.

Ahsan et al. [27] conducted a survey on the subscribers of *Aponjon*, an app that serves pregnant women or new mother. This study concludes that mHealth can significantly contribute to improving health status. Prue et al. [28] found that mobile phone services could assist health professionals by rapidly detecting and treating patients with Malaria in a remote area in Bangladesh. Khatun et al. [29] conducted a survey on 4915 randomly selected personals and proposed a framework to assess the community readiness (in terms of technological, motivational and resource) towards mHealth services in Bangladesh. The study found that the community has access to mHealth services, but a noticeable gap is present between the users' readiness and technological competencies. In another study, Khatun et al. [30] reaffirmed this framework by conducting 37 in-depth interviews with the aim to identify the potential obstacles and possible solutions for mHealth services in Bangladesh. Similarly, another in-depth interview study was conducted by Eckersberger et al. [31] to explore if women are interested to receive text messages on their phone for contraceptive use and family planning in Bangladesh. Huda et al. [32] explored the feasibility, acceptability, and perceived appropriateness of using the mobile phone (in a poor rural community in Bangladesh) to receive a voice message, direct counselling, and unconditional cash transfers to provide food and nutrition guide during pregnancy and the first year of a child's life. The study found that most of the mothers did not feel any major problem to operate mobile phones and the mothers provided very positive feedbacks that support the feasibility, acceptability, and appropriateness of the mobile messaging program.

A few studies also discussed the design, development, evaluation of mobile health applications. For example, Islam et al. [33] discussed the design and development of a mHealth application named *DiaHealth*. They examined the diabetes related health apps and extracted the most important features for developing *DiaHealth* app for diabetes. In another study, Shermin et al. [34] briefly discussed *VirtualEyeDoc* app and its effectiveness. This app is designed in Bengali language to help Bangladeshi people to identify their vision problem and maintain good eye health. Zaman & Mamun [35] designed a preventive app for cardiovascular diseases and depicted the importance of such app in Bangladesh. Hoque et al. [36] developed a mobile-based remote monitoring system to aid the palliative care treatment for rural breast cancer patients in Bangladesh. Khan et al. [37] designed a mHealth app, *PurpleAid* for the women of Bangladesh. The app is designed to diagnose women specific diseases and provide suggestions to prevent and cure.

A number of studies also investigated the adoption and post-adoption of mHealth applications in Bangladesh. For example, Hoque et al. [38] proposed a theoretical model that predicts behavioral intention to adopt mHealth applications of the elder citizens in Bangladesh. They used the Unified Theory of Acceptance and Use (UTAUT) model and derived a set factors of mHealth acceptance in Bangladesh. The factors include gender, the experience of using a mobile phone, performance expectancy, effort expectancy, social influence, facilitating condition, hedonic motivation, price, and habit. Akhter et al. [39] developed a service quality model in the context of mHealth services by framing its relationship with satisfaction, continuance intention and quality of life. In another study, Akhter et al. [40] developed an mHealth continuance model by incorporating the role of service quality and trust.

Finally, only a few studies investigated usability and user experience of mHealth services. Haque et al. [36] developed a mobile application for breast cancer patients in Bangladesh by following a participatory design process involving both patients and doctors. They also explored the key quality of such kind of system to accept and adopted by users of Bangladesh. The study found that the quality of such a system might depend on usability (effectiveness, efficiency, satisfaction) and sustainability. In another study, Bhuiyan et al. [41] proposed an SMS (short message service) based immunization system and then valuated the usability of a SMS alert system for immunization. Two approaches were used to test the usability: (a) *system centric* that measures the availability, localizability, supportability, security, and technical feasibility of the mHealth service; and (b) *human centric* that measures the knowledge of usage, learnability, memorability, and satisfaction of the user. As outcome, this usability study found that the SMS alert system was effective to the end users, found very easy to use, Learn ability, reliability and satisfaction rate was also higher, and the participants were willing to use the proposed system instead of manual system, which indicate the overall efficacy and efficiency of the proposed SMS based immunization system.

Taken together, our literature review indicates that usability has not been the key focus in most prior research conducted in Bangladesh.

## Methods
We followed a 3-stage research approach to achieve the research objectives. *Stage I* aimed to provide the present status of developed mHealth applications and the other two stages (*Stage II* and *Stage III*) aimed to assess the usability of mHealth applications in Bangladesh. These stages were carried out in sequential order. In stage I, all mHealth applications developed for the users/citizens of Bangladesh were investigated to extract data related to their functionalities/features and targeted users. The extracted data were analyzed to provide the present status, as well as create categories of mHealth applications in Bangladesh. In *Stage II*, we carried out expert inspections [42] on a number of selected apps. The set of heuristics to conduct the expert inspection were extracted and synthesized from the existing literature. The expert inspections were carried out by three Human-Computer Interaction (HCI)/usability experts. Each expert evaluated each of the 36 selected applications. Finally, in *Stage III*, a survey was conducted to get feedback scores from users using System Usability Scale (SUS) [43]. We evaluated 9 mHealth applications in this stage. The feedback scores were collected from 30 participants. An overview of the research method is shown in Fig. 1. The detailed discussions of each stage along with the findings are discussed in the following sections.

### Stage I: app review
This stage of research was aimed to investigate the existing mHealth apps in Bangladesh. We conducted the following two sequential steps in this stage.

#### Application search
We performed keywords based search in the popular app stores: Google Play, Apple App Store, Windows Phone Marketplace, and Blackberry App World during the period of March 2018 – April 2018. The keywords are 'Mobile health Bangladesh', 'Apps for healthcare Bangladesh', 'Fitness Bangladesh', 'Doctor Bangladesh', 'Health Tracker Bangladesh', 'Bangladesh Health Doctor', 'Healthy Bangladesh', 'Digital Health Bangladesh', 'Bangla Health Guide', 'Healthy Bangladesh Citizens', 'Pocket Medicine Bangladesh', and 'Bangladesh health care'. Each app appeared in the search results was checked meticulously to ensure whether the application is for health care service and targeted for the users from Bangladesh.

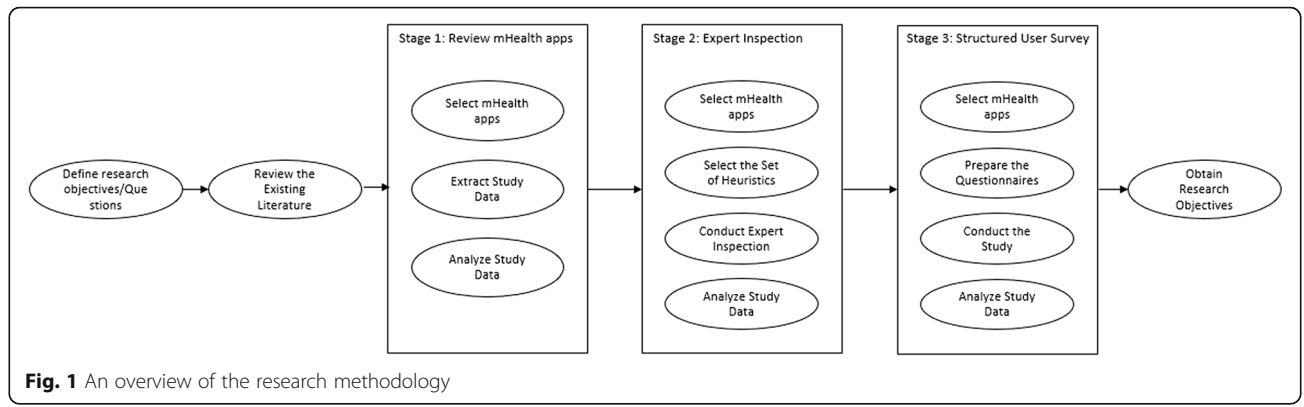

**Fig. 1** An overview of the research methodology

*Analysis and results*
We found 234 mobile health applications that are developed for the users of Bangladesh. These apps were analyzed to understand the present status of mobile health applications in Bangladesh from three perspectives: health services covered by the mobile applications and targeted end users of these applications.

**Clustering mHealth apps** Each app was analyzed to extract information related to its functionalities or features. The functionalities of the mHealth apps were analyzed through an affinity diagram [44]. All four authors of this article participated in the clustering process in two independent expert groups. First, the apps were stack in different or similar cluster based on its functionalities by the two expert groups. Second, the expert groups added cluster headers. Third, the expert groups drew the affinity diagram by organizing these on the board and then expert groups reviewed the clusters and relations between the clusters and then modified the diagram where needed. Finally, these two sets of diagrams were analyzed by all experts to prepare one affinity diagram and come up with the final clusters.

At the end of the process, we found nine clusters: General Health Informative Apps, Institutional Apps, Fitness Apps, Physician Information, Mother & Child, Disease Specific Care App, Food & Nutrition, Herbology, and Homeopathic. The number of applications in each cluster is presented in Table 1. We found that one third of the total apps were developed to provide information related to health care, followed by institutional apps (12%). 10% of the apps were related to physician information whereas, another 10% was related to body fitness.

Similarly, 9% of apps were related to mother & child, whereas, another 9% was related to herbology. Finally, food & nutrition accounted for 7%, disease specific care apps accounted for 5%, and homeopathic accounted for 4% of the total apps.

General Health Informative Apps provide generic information related to the health. Example apps in this category includes Patient Aid, and Bangla Health Guide. Physician Information apps provide information about doctors including their expertise, appointment procedure, contact information, hospital locations, and treatment time. This category includes apps such as Practical Diagnosis and Management, Doctora, and Ms. Daktar. The category, Institutional Apps includes apps that provide information related to a specific hospital and their services, location, contact information, opening time, and emergency number. Example apps in this category include Hashpatal and BD Hospitals. Apps that provide information/assistance for physical exercise, weight loss or gain, and different types of body building exercises belong to the Fitness Apps category. Example apps in this cluster are Bangla Gym Guide, and Yoga Guide Bangladesh. Mother & Child apps were developed to provide all sorts of information and support to pregnant women, and new mother. Examples in this category are Aponjon, and Momota. Applications that provide information (remedies and precautions) about a particular disease belong to the Disease Specific Care Apps category. For example, Breast Screening, and Rabies fall in this category. Food & Nutrition cluster includes apps that provide information about healthy food, and nutrition information for good health. Example apps that fall in this category are Bangla Vitamin Guide, and Pusti Kotha. All herbal treatment, herbal medicine, and herbal information related applications are grouped under the Herbology cluster. Herbology apps also provide all sorts of herbal remedies and their preventive and curative properties. For example, Bangladeshi Herbal Treatment, and Herbal Medicine Bangla apps fall in this category. Finally, apps that provide information and medication support related to the Homeopathic treatment belong to the Homeopathic cluster. Example apps in this category are Homeopathy Guide, and Homeopathic Bangla Book.

**Table 1** Number and percent of apps belonging to different cluster

| Cluster Names | Frequency | Percent |
|---|---|---|
| General Health Informative Apps | 78 | 33 |
| Physician Information | 24 | 10 |
| Institutional Apps | 27 | 12 |
| Fitness Apps | 24 | 10 |
| Mother & Child | 22 | 9 |
| Disease Specific Care Apps | 12 | 5 |
| Herbology | 21 | 9 |
| Food & Nutrition | 17 | 7 |
| Homeopathic | 9 | 4 |

**Targeted users of mHealth apps** We found that the targeted users of mHealth applications were healthcare professionals (4%), mom & kids (15%), patients (24%), and other users (57%) (e.g. health-conscious people). It was striking to observe that none of the apps was developed for the disabled people. We also found that none of the applications were developed for illiterate people. These results indicate that mHealth apps in Bangladesh overlooked marginalized people.

### Stage II: heuristic inspection
In this stage, we randomly selected four mobile applications from each category (i.e., a total 36 apps) for heuristic usability evaluation.

#### Heuristic selection
Prior literature proposed different sets of heuristics to evaluate usability of web and desktop applications such as Nielsen's 10 general principles for interaction design [45], Tognazzini's 16 principles of user interface (UI) design [46], Garrett's 7 guidelines to design websites [47], Schneiderman's 8 Golden Rules of Interface Design [48] and Sollenberger's 10 fundamentals for UI design [49]. Among these, Nielsen's 10 general principles are widely used [50]. Thus, we adapted Nielsen's 10 principles in this study to evaluate usability of the selected mHealth applications. The list of heuristics is presented in Appendix 1.

#### Conducting expert inspection
The selected applications were evaluated based on the heuristics presented in Table 6 in Appendix 1 by three usability experts (first three authors of this article). Among them, one had 10 years of experience in UI design and evaluation with a PhD degree in HCI, while other two were postgraduate students (one in PhD and one in MSc program) in Computer Science and had 2–3 years of experience in usability evaluation and UI design. Each usability expert conducted the inspection independently. Individual evaluation report for each application was then compared to write a single report for each application. In some cases, there were conflicts raised about the revealed problems. These conflicts were solved by discussions. For heuristic evaluation, Nielsen's (1995) [45] concept of severity rating (0 to 4) was adopted, with *0*, indicating not a usability problem at all; *1*, cosmetic problem only; *2*, minor usability problem; *3*, major usability problem; and *4*, usability catastrophe. A few examples of usability problems for *CareSatisfaction* application are presented in Appendix 2.

### Analysis and results
The study data was analyzed to find out the overall usability, heuristics that are violated, and severity associated with each heuristic violation. We found 406 problems by evaluating 36 mHealth applications with an average severity of 2.74 (see Table 2). Around 21% of occurrences were related to the violation of the heuristic 'Aesthetic and minimalist design (H7)'. The least (around 4%) violated heuristic was the 'Customization and shortcuts (H6)'. In general, heuristics related to the user control and freedom (H2), Consistency and Standards (H3), and readability and glanceability (H10) showed high severity while physical interaction and ergonomics (H8) showed less severity. More than half (61%) of the total revealed problems were Catastrophic and Major problems. Catastrophic problems were observed in terms of all heuristics, but maximum number is found for H7 and H10.

Apps related to the Physician Information, Institutional Apps, and Disease Specific Care Apps showed highest percent of usability problems, while the apps related to Fitness Apps and Herbology showed highest severity score (see Table 3). Slightly smaller numbers of usability problems were found for apps that belong to Mother & Child and Food & Nutrition cluster. The average severity level varies from 2.1 to 3, with Herbology showing the maximum severity score. A high number (from 16 to 33%) of catastrophic problems were found for each cluster.

Table 4 shows the number of times each of the heuristics was violated. As shown in the table, each of the heuristics was violated from minimum one to maximum 12 times for each cluster except the Mother & Child cluster.

We calculated the number of usability problems and the number of UIs in each application. The Pearson

**Table 2** Number of unique usability problems identified by the evaluators

| Heuristics | Severity | | | | Total | % of Problems | Avg Severity |
|---|---|---|---|---|---|---|---|
| | Cosmetic | Minor | Major | Catastrophic | | | |
| H1: Visibility of system status | 7 | 9 | 16 | 7 | 39 | 10 | 2.6 |
| H2: User control and freedom in system navigation | 4 | 12 | 19 | 15 | 50 | 12 | 3.1 |
| H3: Consistency and standards | 9 | 5 | 23 | 11 | 48 | 12 | 3.1 |
| H4: Realistic error management | 9 | 13 | 7 | 4 | 33 | 8 | 2.6 |
| H5: Minimize the user's memory load | 3 | 7 | 14 | 5 | 29 | 7 | 2.7 |
| H6: Customization and shortcuts | 2 | 5 | 7 | 4 | 18 | 4 | 2.7 |
| H7: Aesthetic and minimalist design | 13 | 15 | 33 | 23 | 84 | 21 | 2.5 |
| H8: Physical interaction and ergonomics | 5 | 11 | 4 | 3 | 23 | 6 | 2.4 |
| H9: Minimize human-device interaction | 5 | 7 | 14 | 4 | 30 | 7 | 2.5 |
| H10: Readability and Glanceability | 7 | 10 | 12 | 23 | 52 | 13 | 3 |
| Total | 64 | 94 | 149 | 99 | 406 | 100 | 2.74 |

Correlation Coefficient tests between the number of UIs and the problems observed ($r = 0.0899$, $p = 0.656$) and between the number usability problems and the average severity rating of each application ($r = 0.3128$, $p = 0.112$) were non-significant.

## Stage III: user study

### Participants profile

We prepared a list of 50 individuals who could participate in our study using snowball sampling. We sent email invitations to these individuals for participating in our study. Among them 30 individuals agreed to participate in this study. Therefore, a structured user survey was conducted among these 30 (17 males and 13 females) participants. The recruited participants' profession includes undergraduate students, banker, teacher, intern doctor and housewife. The participants' average age was approximately 33 years and ranged from 23 to 55 years. All the participants are experienced with smart phones and mobile applications. Twenty-one participants had experience with the use of mHealth applications and used 2–3 mobile health applications.

### Study procedure

The study was conducted in a Software Engineering lab at the authors' institute following the System Usability Scale (SUS) [43] evaluation procedure. The SUS is a simple, ten items scale that provides the overall view of subjective usability assessment. The SUS consists of 10 items, with odd-numbered items worded positively and even-numbered items worded negatively. The questions are included in SUS are presented in Appendix 3. SUS is a Likert scale, and the respondents indicate their level of agreement or disagreement on a scale of 1 to 5 for each statement. Nine mHealth apps, 1 from each category were selected for evaluation. The applications that had maximum downloads were selected for user testing. The test was conducted in three sessions, where each app was evaluated by 10 participants (i.e., each participant evaluated 3 apps selected in a random fashion). In each session, one participant evaluated 1 app at a time and took on average 25–30 min. During each test-session, at first the participant was briefed about the purpose of the study and his/her roles. The participant was also briefed that the goal of the study is not to assess him/her, rather to evaluate the mobile application, so that he/she behaves normally during the test session and provides honest opinion/response about the overall system functionalities and performance. After that, the test participant's demographic information was collected, and a test-consent form was also signed with him/her. Then the participant was asked to explore and use a randomly assigned application for 15–20 min. In the end, the participant was asked to answer the SUS statements [43] . The SUS score was calculated following the Brooke's (1996) [43] guidelines. First, for odd numbered questions, we subtracted one from the user response (scale position), and for even-numbered questions, we subtracted the user responses (scale position) from 5. After that, we summed up the converted score and then multiplied it by 2.5. Finally, we computed the average value of the SUS scores for the studied applications.

**Table 3** Number of usability problems to each cluster

| Cluster | Severity | | | | Total | % of Problems | Avg Severity |
|---|---|---|---|---|---|---|---|
| | Cosmetic | Minor | Major | Catastrophic | | | |
| General Health Informative Apps | 8 | 12 | 12 | 12 | 44 | 11 | 2.75 |
| Physician Information | 7 | 9 | 21 | 18 | 55 | 14 | 2.8 |
| Institutional Apps | 6 | 12 | 23 | 13 | 54 | 13 | 2.65 |
| Fitness Apps | 9 | 17 | 14 | 8 | 48 | 12 | 3 |
| Mother & Child | 7 | 12 | 7 | 5 | 31 | 8 | 2.1 |
| Disease Specific Care Apps | 7 | 10 | 27 | 10 | 54 | 13 | 2.8 |
| Herbology | 8 | 4 | 18 | 12 | 42 | 10 | 3 |
| Food & Nutrition | 5 | 7 | 16 | 9 | 37 | 9 | 2.75 |
| Homeopathic | 7 | 11 | 11 | 12 | 41 | 10 | 2.6 |
| Total | 64 | 94 | 149 | 99 | 406 | 100 | 2.74 |

**Table 4** Heuristics violation to each cluster

| Cluster | Heuristics | | | | | | | | | | Total |
|---|---|---|---|---|---|---|---|---|---|---|---|
| | H1 | H2 | H3 | H4 | H5 | H6 | H7 | H8 | H9 | H10 | |
| General Health Informative Apps | 4 | 6 | 5 | 4 | 2 | 4 | 12 | 3 | 3 | 1 | 44 |
| Physician Information | 7 | 7 | 6 | 4 | 5 | 2 | 10 | 2 | 4 | 8 | 55 |
| Institutional Apps | 6 | 7 | 7 | 3 | 5 | 3 | 9 | 4 | 3 | 7 | 54 |
| Fitness Apps | 4 | 7 | 4 | 4 | 3 | 1 | 11 | 3 | 4 | 7 | 48 |
| Mother & Child | 4 | 5 | 4 | 1 | 2 | 0 | 8 | 1 | 2 | 4 | 31 |
| Disease Specific Care Apps | 4 | 5 | 5 | 5 | 4 | 3 | 12 | 3 | 5 | 8 | 54 |
| Herbology | 3 | 4 | 7 | 4 | 3 | 2 | 9 | 2 | 3 | 5 | 42 |
| Food & Nutrition | 4 | 3 | 4 | 5 | 2 | 2 | 7 | 2 | 2 | 6 | 37 |
| Homeopathic | 3 | 6 | 6 | 3 | 3 | 1 | 6 | 3 | 4 | 6 | 41 |
| Total | 39 | 50 | 48 | 33 | 29 | 18 | 84 | 23 | 30 | 52 | 406 |

### Data analysis and outcomes

The SUS score of each app is shown in Table 5. We followed Bagor et al.'s (2008) [51] guidelines to evaluate the SUS scores. The score can be acceptable (SUS score > 70), marginal (50 < SUS score > 70) and unacceptable (SUS score < 50). The results show that more than 50% of the application are below the acceptable score. Among the accepted applications, most of the apps were found to have SUS scores just above the threshold value of acceptable. Only *Aponjon Pregnancy* was found to have high SUS evaluation score. Taken together, most of these apps were found to have poor to marginal usability according to their SUS scores.

**Table 5** SUS score of the studied applications

| Cluster | App name | Average SUS Score | Remarks |
|---|---|---|---|
| General Health Informative Apps | Patient Aid | 65.7 | High-marginal |
| Physician Information | DIMS | 73.2 | Acceptable |
| Institutional Apps | Hospital Finder | 52.8 | Low-marginal |
| Fitness Apps | Bangla Gym Guide | 70.5 | Acceptable |
| Mother & Child | Aponjon Pregnancy | 79.2 | Acceptable |
| Disease Specific Care Apps | এলার্জিৰ সহজ চিকিৎসা <Easy Treatment for Allergy> | 69.3 | High-marginal |
| Herbology | ভেষজ চিকিৎসা <Herbal Treatment> | 68.3 | High-marginal |
| Food & Nutrition | Fruits Benefit in Bangla | 71.7 | Acceptable |
| Homeopathic | Homeopathic Bangla book | 65.4 | High-marginal |

Figure 2 represents the relation between the average SUS scores and the number of usability problems found. The results show that SUS scores are inversely related to the average number of problems found in *Stage II* (see Fig. 2). This indicates that the findings of expert inspection and SUS evaluation match with each other in most cases. For those apps where we observed less number of usability problems (7–8 problems), we found the SUS score values were acceptable. Those apps with comparatively higher number of usability problems revealed SUS score below the acceptable level. However, exception was found for two apps (*Herbal Treatment* and *Homeopathic Bangla Book*). Although, the number of usability problems of these apps was comparatively smaller, their SUS scores were below the acceptable level. We also investigated if SUS scores relate to application size (number of UIs). The result is shown in Fig. 3. We found that SUS scores do not relate to the number of UIs.

## Discussions
### Key findings
In this paper, we investigated the overall usability of mobile health applications in Bangladesh through an extensive empirical research. This research provides three main findings. First, we summarized an overview of

mHealth applications developed in Bangladesh along with the targeted end users and the features/functionalities provided by the applications. We found 9 categories of applications based on their features/functionalities. The study also found that healthcare professionals, mothers, patients, and other general users have been the targeted end users for the mHealth apps in Bangladesh.

Second, the study found that the usability of selected mHealth applications in Bangladesh in general does not adhere to or follow the design principles. Both expert inspection and SUS technique revealed that the existing mHealth applications would be very hard for the users to use. The results indicate that the lack of usability would be one of the key barriers for adopting mHealth applications in Bangladesh.

Third, we also strived to explore the possible relation among the number of usability problems with the application size (number of UIs), the severity scores, and the SUS score. We found that although positive relations were observed between the number of usability problems with the number of UIs and with the severity score, but the relations were not significant. Furthermore, we found that the number of usability problems had an inverse relation with the SUS score. This indicates that the findings from heuristic inspection and SUS evaluation complement each other.

However, a major limitation of this study is that we conducted the user study with a small number of participants and with a small number of apps. Another limitation is that we have not compared between Government and private apps in this study.

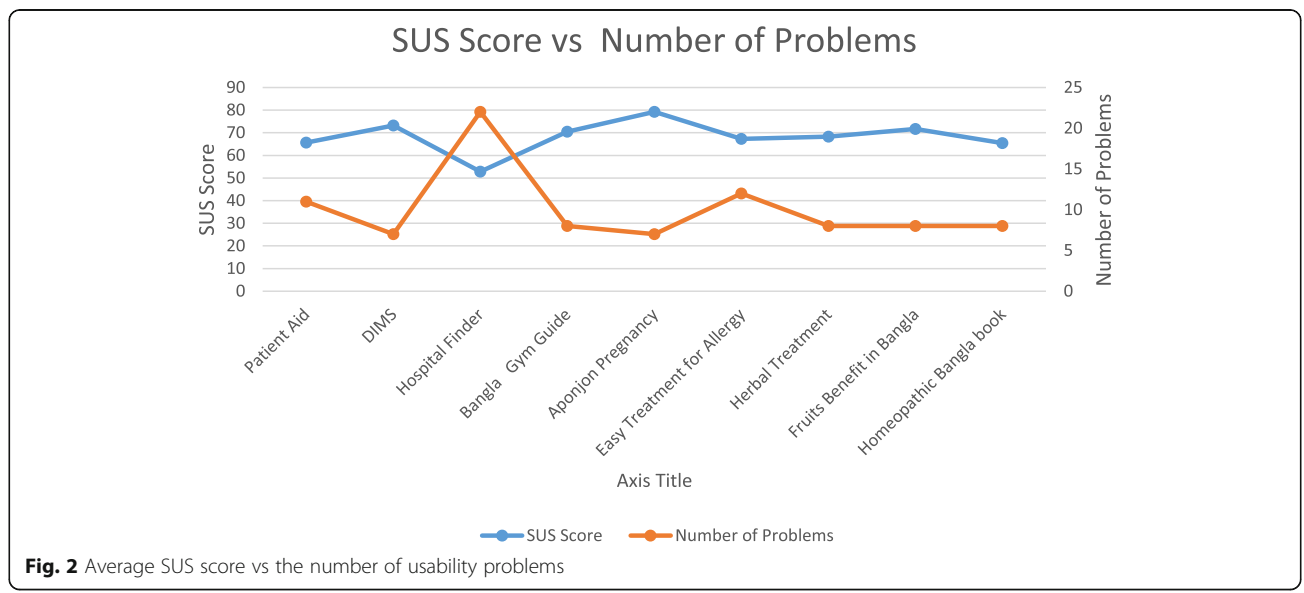

**Fig. 2** Average SUS score vs the number of usability problems

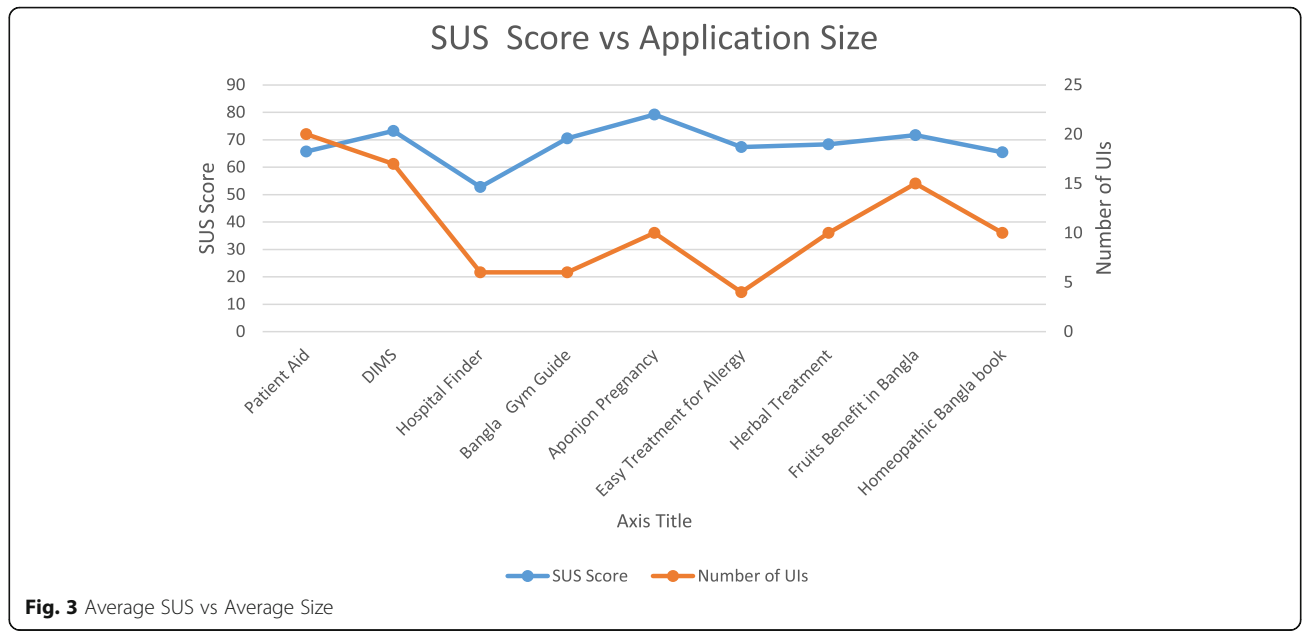

**Fig. 3** Average SUS vs Average Size

**Implications and directions for future research**

This research is the first to provide in-depth view on the usability status of mHealth applications in Bangladesh. The findings will greatly contribute to research in health informatics, and to the HCI practitioners, non-government organizations (NGOs) and Government organizations providing health services in Bangladesh. The study provides the following implications: First, our results provide clear empirical evidence that usability has been overlooked in mHealth applications development. These findings could provide important practical implications such as increasing awareness of improving usability in mHealth applications. The results also depicted the type of usability problems (or heuristics violated in development), found in the tested apps. These findings may provide the underlying knowledge (e.g., heuristics often not fulfilled) for usable mHealth applications development. Furthermore, the findings also reveal the necessity of curriculum revision in computer science education to integrate HCI concepts. Consequently, possible future research area could be to investigate the underlying reasons and its possible solutions of why usability was overlooked in the developed mHealth applications.

Second, the findings of this research can guide researchers for further studying mHealth services and plan for usability-based research designs in the context of Bangladesh. The lack of usability in tested apps imposes an important question of to what extent user centered development approaches are actually used in Bangladesh for developing the mHealth applications. Furthermore, it raises the question of to what extent requirements elicitation study for finding the contextual need for developing new and usable mHealth applications has been conducted in developing these applications. All these questions provide further research directions that can be conducted by researchers. Future research, for example, may focus to design, develop and evaluate mHealth applications by following methodologies that keep usability at the center of attention.

Third, this paper is the first to provide a taxonomy of mHealth apps developed in Bangladesh. We found 9 types of mHealth applications available in Bangladesh. Future studies can be conducted to verify the applicability of this taxonomy in other contexts and possibly extend it.

Finally, our study results suggest that there have not been enough effort to develop mHealth apps for marginalized population in Bangladesh. Thus, special emphasis on user profile (such as disability, literacy, etc.) would be required to develop accessible and usable apps for the marginalized citizens of Bangladesh.

**Conclusions**

The purpose of this study was to explore usability of mHealth applications in Bangladesh. The findings suggest that the usability of the mHealth applications is not satisfactory. Our findings highlight the heuristics that are often violated in these mHealth applications. Such information can be used by designers to improve their applications in terms of usability. This in turn, may impose a barrier on adoption and use of mHealth applications in Bangladesh.

## Appendix 1
### Heuristics of Expert Inspection

**Table 6** Set of heuristics for expert inspection

| No | Heuristics |
|---|---|
| H1 | Visibility of system status |
| | a. Always keep users informed about what is going on (e.g. "loading", "deleted") |
| | b. Provide appropriate feedback (e.g. tactile, visual, audible) to the user within reasonable time |
| H2 | User control and freedom in system navigation |
| | a. Provide "emergency exits" to instantly leave an unwanted state |
| | b. Provide basic navigation controls on screen, even if the device itself provide buttons to perform similar functions (e.g. "back" function) |
| | c. Avoid accidental activation of closely located touch controls |
| H3 | Consistency and standards |
| | a. Users should be able to do things in a familiar, standard and consistent way, based on experience from similar apps and platforms (e.g. tactile gestures) |
| H4 | Realistic error management |
| | b. Express error in plain language (no codes) and constructively suggest a solution |
| | c. Users should be warned to confirm risky action in order to avoid accidental errors (e.g. delete, payment) |
| H5 | Minimize the user's memory load |
| | a. Provide clear affordance for touch controls and other UI elements |
| | b. User should not have to remember information between screens |
| H6 | Customization and shortcuts |
| | a. Make the system easy for first time setting and learning |
| | b. Provide basic configuration options for common users and advanced configuration for expert users |
| | c. Provide shortcuts to most frequent tasks, allow users to tailor frequent actions. |
| H7 | Aesthetic and minimalist design |
| | a. Avoid displaying irrelevant or rarely needed information |
| | b. Most relevant information should be highlighted through large size, color, etc. |
| H8 | Physical interaction and ergonomics |
| | a. Touch control elements should have adequate size and spacing for fat fingers |
| | b. Place control elements in a recognizable position |
| | c. Place control elements so they can be easily pressed with the user's thumb in any hand (or provide an option for switching layout based on hand orientation) |
| H9 | Minimize human-device interaction |
| | a. Strive to reduce interaction effort because users may be in motion with only one-thumb and one-eye on the system |
| | b. Reduce data entry, especially with typing. Use sensors (e.g. location, voice) or historical and personalization data to establish defaults |
| H10 | Readability and Glanceability |
| | a. Ensure that text and textboxes fit on the screen |
| | b. Ensure readability in different lighting conditions with sufficient contrast |
| | c. Ensure user is able to quickly get relevant information by glancing at screen |

## Appendix 2

### Evaluation of *CareSatisfaction* App

A few examples of problems by an evaluator for *CareSatisfaction* application (see Fig. 2) are presented in Table 3. The *CareSatisfaction* app provides patients opinions about the service quality of clinics or hospitals in Bangladesh.

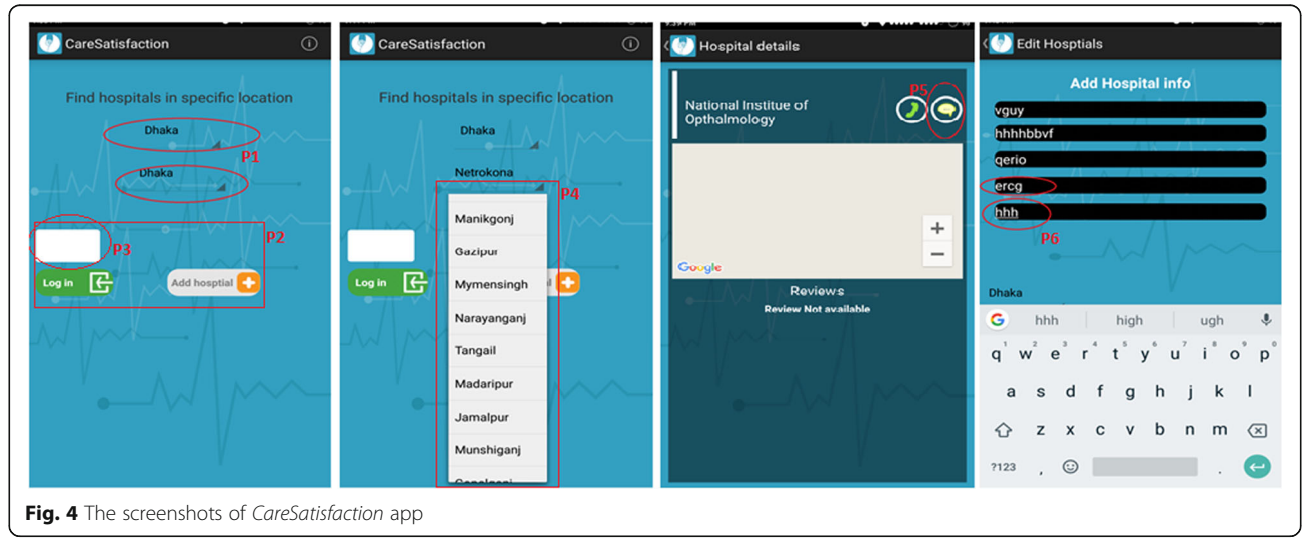

**Fig. 4** The screenshots of *CareSatisfaction* app

**Table 7** Example of predicted problems in *CareSatisfaction* app

| Problem No (see Fig. 2) | Description (comment from the evaluator) | Severity | Violated Heuristic |
|---|---|---|---|
| P1 | "The hint provided about the text input is confusing since in both fields same example texts are shown by default. The first field should be for the Division whiles the second one for the District name. So, user will not be able to quickly grasp the information or what they need to do" | 3 | H10 |
| P2 | "The orientations of the three buttons are very unusual. Despite of having space at left and below, the buttons are organized in a very compact way. The buttons are not organized in a conventional manner and lack proper spacing among the buttons" | 2 | H8, H9 |
| P3 | "No proper instruction, indication or hints is given about the button. Even it is almost impossible to detect it as a button" | 4 | H3, H10 |
| P4 | "The lists are organized in a scattered way. Difficult for a user to find the certain area/district. A systematic way could be followed, for example, ordering the districts name alphabetically. User has no control to see the list in a systematic way" | 2 | H4, H6, H9 |
| P5 | "The message icon is used instead of a rating icon for asking user to rate the app. It may reduce the users' memory load (user should not remember its functionality) if a proper icon was used here" | 3 | H3, H5 |
| P6 | "No constraint as well as error control mechanism are given in the fields of adding new hospital. For example, user can give alphabetical input in latitude longitude field (where numerical value could be the only accepted value)" | 2 | H4 |
| P7 | "Basic navigation control is missing in the screen like there is no Home or Back button is present in the screen" | 3 | H2 |

# Appendix 3
## List of SUS Questions

Q1- I think that I would like to use this system frequently.

Q2- I found the system unnecessarily complex.

Q3- I thought the system was easy to use.

Q4- I think that I would need the support of a technical person to be able to use this system.

Q5- I found the various functions in this system were well integrated.

Q6- I thought there was too much inconsistency in this system.

Q7- I would imagine that most people would learn to use this system very quickly.

Q8- I found the system very cumbersome to use.

Q9- I felt very confident using the system.

Q10- I needed to learn a lot of things before I could get going with this system.

## Abbreviations
HCI: Human-computer interaction; HIV: Human immunodeficiency virus; ICT: Information and communication technology; MAMA: Mobile alliance for maternal action; MEASURE: Monitoring and evaluation to assess and use results; NGO: Non-Government Organization; PDA: Personal digital assistant; SMS: Short message service; SUS: System usability scale; UI: User interface; US: United States; USAID: United States Agency for International Development; UTAUT: Unified theory of acceptance and use

## Acknowledgements
The authors would like to thank the participants whose participation made the study possible. Their efforts are gratefully acknowledged.

## Authors' contributions
All authors contributed to write the article, while MNI contributed as a lead author for defining the problem statement, conducting the studies, analyzing the data and wrote a major portion of the article. The app review, expert inspection and user survey were carried out by the MNI, MMK, and TTI, while data analysis was performed by MNI. AKMNI contributed toward rewriting the entire draft article to prepare it for publication. All authors read, edited, and approved the final manuscript.

## Funding
Not Applicable.

## Availability of data and materials
The datasets used and/or analysed during the current study are available from the corresponding author on reasonable request.

## Ethics approval and consent to participate
We confirm that ethical approval was applied for conducting this research. The target of the study was to collect minimal data from human participants on the subjective usability assessment of the selected mobile applications. No human data, human tissue or any clinical data were collected for this study. Therefore, the ethical committee headed by the Research & Development Wing of Military Institute of Science and Technology (MIST) decided that it is not required to have formal approval. We also declare that we have taken written consent from study III participants to participate in the evaluation study.

## Consent for publication
We have not used images or other personal or clinical details of participants that compromise anonymity. We declare that we have collected written consent from study III participants that the data collected from them will contribute to research publications.

## Competing interests
The authors declare that they have no competing interests.

## Author details
$^{1}$Department of Computer Science and Engineering, Military Institute of Science and Technology, Dhaka 1216, Bangladesh. $^{2}$Department of Future Technologies, University of Turku, Turku, Finland.